\documentclass[fleqn,usenatbib]{mnras}

\usepackage{newtxtext,newtxmath}

\usepackage[T1]{fontenc}
\usepackage{ae,aecompl}

\usepackage{graphicx}	
\usepackage{amsmath}	
\usepackage{amssymb}	

\title[NGC 1052: stellar population]{A panchromatic spatially resolved study of the inner 500\,pc of NGC\,1052 - I: Stellar population}
\author[Dahmer-Hahn et al.]{L. G. Dahmer-Hahn $^{1}$\thanks{E-mail: dahmer.hahn@ufrgs.br}, R. Riffel$^1$, J. E. Steiner$^2$, R. A. Riffel$^3$, R. B. Menezes$^4$, \newauthor  T.V. Ricci$^5$, N.Z. Dametto$^1$,  T. Storchi-Bergmann$^1$, M. R. Diniz$^3$
\\
$^{1}$Departamento de Astronomia, Universidade Federal do Rio Grande do Sul. Av. Bento Goncalves 9500, 91501-970, Porto Alegre, RS, Brazil.\\
$^{2}$Instituto de Astronomia, Geof\'isica e Ci\^encias Atmosf\'ericas, Universidade de S\~ao Paulo, 05508-900, S\~ao Paulo, Brazil\\
$^{3}$Universidade Federal de Santa Maria, Departamento de F\'isica, Centro de Ci\^encias Naturais e Exatas, 97105-900, Santa Maria, RS, Brazil\\
$^{4}$Centro de Ci\^encias Naturais e Humanas, Universidade Federal do ABC, 09210-580, SP, Brazil\\
$^{5}$Universidade Federal da Fronteira Sul, Campus Cerro Largo, RS 97900-000, Brazil
}

\date{Accepted XXX. Received YYY; in original form ZZZ}

\pubyear{2015}

\begin{document}
\label{firstpage}
\pagerange{\pageref{firstpage}--\pageref{lastpage}}
\maketitle

\begin{abstract}
We map optical and near-infrared (NIR) stellar population properties of the inner 320$\times$535\,pc$^2$ of the elliptical galaxy NGC\,1052. The optical and NIR spectra were obtained using the Gemini Integral Field Units of the GMOS instrument and NIFS, respectively. By performing stellar population synthesis in the optical alone, we find that this region of the galaxy is dominated by old (t$>$10\,Gyr) stellar populations. Using the NIR, we find the nucleus to be dominated by old stellar populations, and a circumnuclear ring with younger ($\sim$2.5\,Gyr) stars.  We also combined the optical and NIR datacubes and performed a panchromatic spatially resolved stellar population synthesis, which resulted in a dominance of older stellar populations, in agreement with optical results. We argue that the technique of combining optical and NIR data might be useful to isolate the contribution of stellar population ages with strong NIR absorption bands. We also derive the stellar kinematics and find that the stellar motions are dominated by a high ($\sim$240km$\cdot$s$^{-1}$) velocity dispersion in the nucleus, with stars also rotating around the center. Lastly, we measure the absorption bands, both in the optical and in the NIR, and find a nuclear drop in their equivalent widths. The favored explanation for this drop is a featureless continuum emission from the low luminosity active galactic nucleus.
\end{abstract}




\section{Introduction}
In order to access the stellar population content in galaxies, the technique of stellar population synthesis \citep[e.g. ][]{bica88,CF04,CF05,FIREFLY} has been widely used over the years, with many key results in galaxy evolution being obtained \citep{CF04,CF05,rembold+17,zheng+17,goddard+17a,goddard+17b,deAmorim+17,mallmann+18}. This stellar population synthesis technique has been employed mostly in the optical spectral region. However, with the development of new telescopes focused in other wavelength ranges, it became clear that new evolutionary population synthesis (EPS) models which cover wavelength ranges beyond the optical would have to be developed in order to explore the full potential these telescopes have to offer \citep{M05,riffel+08,noel+13,zibetti+13,MG15,baldwin+18,luisgdh+18}.
\par
Although the use of near infrared (NIR) fingerprints on the study of stellar populations started nearly 40 years ago \citep{rieke+80,origlia+93,oliva+95,engelbracht+98,lancon+01}, only in the last decade there was an improvement on the usage of the whole spectral range \citep{riffel+08,riffel+09,cesetti+09,riffelRA+11,storchi-bergmann+12,kotilanen+12,dametto+14,diniz+17,riffelRA+17}.
Part of this usage has been due to the lower effect of dust reddening in the NIR, meaning that it is possible to access populations in dustier regions, which in the optical and ultraviolet would be obscured. Also, the most prominent spectral features of stars beyond the main sequence (especially thermally pulsing asymptotic giant branch stars, hereafter TP-AGB stars) are located in the NIR \citep{M05,riffel+09,noel+13,zibetti+13,riffel+15}. Thus, the NIR is crucial in order to identify stellar populations which are dominated by these stars.
\par
Two recent papers studied the ability to identify the stellar populations in the NIR compared to the optical region. \citet{baldwin+18}, by studying 12 early-type galaxies (ETG), found that the derived star formation histories vary dramatically in the NIR, if compared to the optical, depending on the chosen stellar spectral library. Also, they found that models based on high-quality spectral libraries fit NIR data better and also produce more consistent results when compared to optical ones, with the ages being imprinted on the shape of the continuum in the case of low spectral resolution models. Also, according to \citet{luisgdh+18}, NIR models with low ($\lesssim$300) spectral resolution have limitations to distinguish between the different stellar populations, since with these models too much weight is given to the shape of the continuum, when deriving the ages of the stellar populations. Combined, these results show that newer models with higher spectral resolution \citep[e.g.][]{MG15,MIUSCATIR, conroy+18} are essential in order to produce more consistent results in the NIR.
\par
\begin{figure}
	\includegraphics[width=\columnwidth]{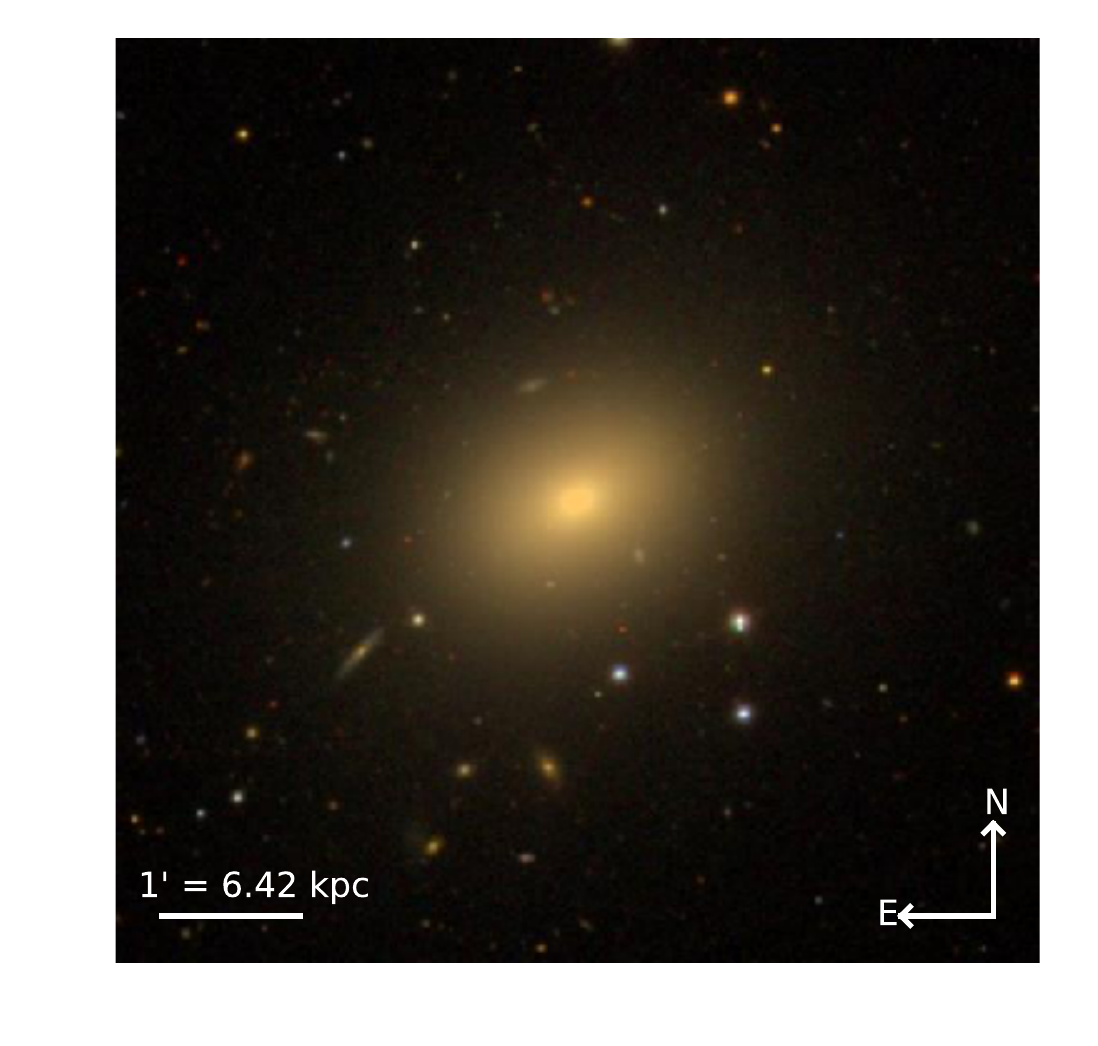}
	\caption{Optical image of NGC\,1052 taken with Sloan Digital Sky Survey \citep{adelman-mccarthy+06,baillard+11}.}
	\label{fig:fig1}
\end{figure}

In this work, by using the new library of models of \citet{e-miles}, which has high ($\sim$2000) spectral resolution both in the optical and in the NIR, we present a study of the stellar population of the nuclear region of NGC\,1052 (Figure~\ref{fig:fig1}). 

This galaxy is a giant elliptical classified as E4 \citep{forbes+01,xilouris+04} and with a redshift of 0.005037, harboring one of the nearest radio-loud active galactic nuclei (AGN). On radio wavelengths, it displays two jets with slightly different orientations \citep{wrobel84,FeyCharlot97}. In X-rays, it shows a compact core, best fitted by an absorbed power law and various jet-related emissions and an extended region \citep{kadler+04}. 
\par
In the optical, its nucleus is classified as low-ionization nuclear emission-line region \citep[LINER,][]{heckman80,ho+97}. \citet{barth+99} used polarized light, and confirmed a hidden BLR emission in the H$\alpha$ line, making NGC 1052 the first LINER to have a broad emission component detected in polarized light. Later, \citet{sugai+05}, using integral field spectroscopic data obtained with the Subaru Telescope at a spatial resolution of $\sim$0.4'', reported the detection of weaker features at the nucleus, including the [Fe\,III] and He\,II emission lines, as well as an unresolved broad component of the H$\beta$ emission. They also found that the spatial structure and velocity field of this source requires the existence of three main components: i) a high-velocity bipolar outflow, ii) low-velocity disk rotation, and iii) a spatially unresolved nuclear component.
\par
Regarding the stellar content of NGC 1052, \citet{raimann+01} used long slit spectra and found that the inner 1\,kpc is dominated by an older and more metallic stellar population ($\sim$10\,Gyr) whereas a younger, 1\,Gyr population becomes increasingly more important outwards. Also, \citet{milone+07} found, by measuring Lick indices, that the bulge of NGC\,1052 is older (12-15\,Gyr) and more metallic than the rest of the galaxy, suggesting this galaxy was formed by processes in which the star formation occurred first at the bulge on short timescales. They also found that, along the major axis, a younger population is responsible for $\sim$30\% of the light fraction at 1.5\,kpc, whereas along the minor axis this stellar population is not important. On the other hand, \citet{pierce+05} used Keck spectra of 16 globular clusters (GC) and a long-slit spectrum of the whole galaxy, finding that its nucleus has a luminosity-weighted central age of $\sim$2\,Gyr and [Fe/H]$\sim$+0.6\,dex, which is consistent with the merger event that occurred $\sim$1\,Gyr ago \citep{gorkom+86}. They also found that the GCs of NGC\,1052 are dominated by a $\sim$13\,Gyr stellar population, with a few of these GCs having strong blue horizontal branches which can not be fully accounted for using stellar population models. Based on the age of the GCs, they argued that, despite the recent merger event, little recent star formation occurred.
\par
Also, \citet{ontiveros+11} found 15 compact sources exhibiting H$\alpha$ luminosities an order of magnitude above the estimated for an evolved population of extreme horizontal branch stars. Their H$\alpha$ equivalent widths and optical-to-NIR spectral energy distributions are consistent with them being young stellar clusters with ages <7\,Myr. According to the authors, this is probably related with the merger event experienced by the galaxy.
\par
Regarding the stellar kinematics, \citet{dopita+15} found that the stars rotate smoothly around the photometric minor axis and are characterized by a velocity dispersion of $\sim$200km$\cdot$s$^{-1}$. They also found that there is a sharp cusp in the velocity dispersion close to the nucleus, presumably within the zone of influence of the central black hole. These results were later confirmed by \citet{riffelRA+17} using NIR data, which also found a centrally peaked $\sigma$ distribution and a rotation around the photometric minor axis. Besides that, they found an anti-correlation between the h$_3$ momentum and the velocity field, which they interpreted as due do the contribution of stars rotating slower than those in the galaxy disk, probably due to motion in the galaxy bulge.
\par
Most modern libraries of simple stellar populations (SSPs) still lack models younger than 1.0\,Gyr in the NIR. This is caused by the low amount of observed hot stars with adequate signal-to-noise ratio and moderate (R>1000) spectral resolution in the NIR. The only library that includes models younger than 1\,Gyr is the E-MILES \citep{e-miles} library. However, as pointed out by the authors, models younger than 0.5\,Gyr are unsafe.
\par
The previous stellar population studies of NGC\,1052, which showed that the inner 1\,kpc is dominated by stellar populations older than 2\,Gyr, make this galaxy an adequate object to have its stellar population properties studied with these modern NIR libraries. Also, while the optical range is more sensitive to bluer stars, the NIR is dominated by the red and cold stars (including the TP-AGB). Combining these two wavelength ranges provides a unique opportunity to test modern stellar population models simultaneously in the optical and in the NIR. Lastly, because the ionization source of the LINER-like emission in NGC\,1052 is still under debate \citep{fosbury+78,fosbury+81,diaz+85,SugaiMalkan00,gabel+00,dopita+15}, combining these two wavelength ranges provides a unique opportunity to search for any contributions of an AGN, since hot dust emission peaks in the NIR and the featureless contribution from the AGN is stronger in bluer wavelengths.
\par
In this work, we aim to study the stellar population properties of NGC\,1052, in the optical and NIR separately, and also combining both wavelength ranges in a single fit. This paper is structured as follows: the data and reduction processes are presented in Section~\ref{sec:data}. In Section~\ref{sec:synth} , we present the method of spectral synthesis used throughout the paper. The results are presented in Section~\ref{sec:results}, and are discussed in Section~\ref{sec:discussion}. Lastly, the conclusions are drawn in Section~\ref{sec:conclusions}.

\section{Data And Reduction}
\label{sec:data}

In order to fully explore the inner region of NGC\,1052, we used two sets of datacubes, one in the optical and the other in the NIR. The optical datacube was obtained with Gemini Multi-Object Spectrograph \citep[hereafter GMOS,][]{hook+04, allington-smith+02} and the NIR one was obtained with the Near-infrared Integral Field Spectrograph \citep[hereafter NIFS,][]{nifs}, both instruments attached to the Gemini telescopes. Below we describe the observations and data reduction process. 

\subsection{Optical data}
The optical data were obtained on September 30th 2013 as part of GS-2013B-Q-20 Gemini South project using GMOS on Integral Field Unit (IFU) mode. The original Field of View (FoV) is 3\farcs5$\times$5\farcs0 (320$\times$535\,pc$^2$), with a natural seeing of 0\farcs88, estimated from stars present in the acquisition image, which was taken immediately before the datacube observation. One 30 min exposure was taken, using the B600-G5323 grating, with a central wavelength of 5620 \AA. The FWHM is equal to 1.8\,\AA$ $ throughout the spectra.

\par

The data were reduced using the \textsc{Gemini/IRAF} package and included: trimming, bias subtraction, bad pixel removal, cosmic ray removal \citep[using the L. A. Cosmic routine][]{vandokkum01}, extraction of the spectra, GCAL/twilight flat correction, wavelength calibration, sky subtraction and flux calibration (taking into account the atmospheric extinction). Finally, the datacube was constructed, with square spatial pixels (spaxels) of 0\farcs05 width.

\par

\begin{figure}
	\includegraphics[width=\columnwidth]{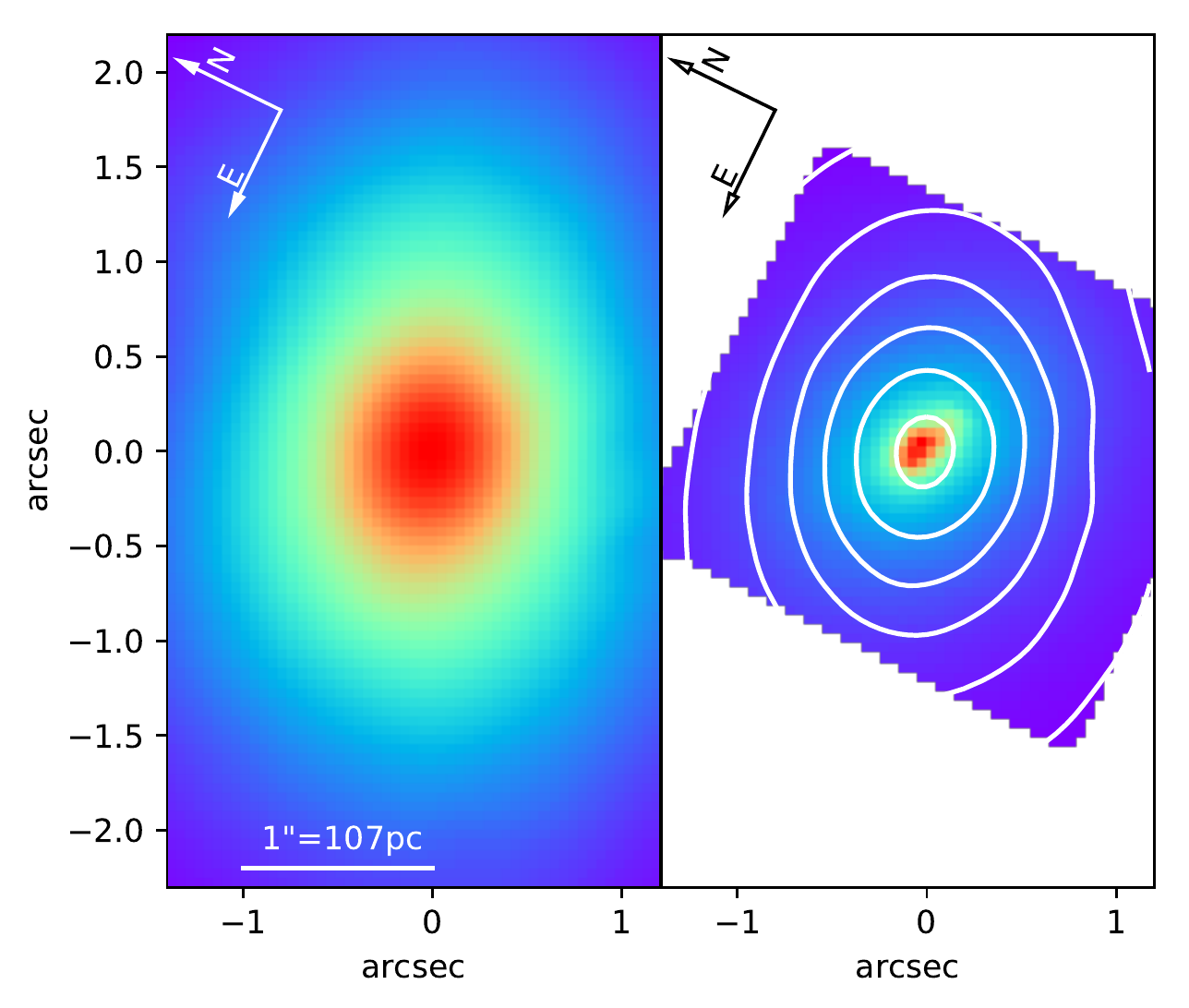}
	\caption{Left: optical continuum of NGC\,1052 obtained with GMOS. Right: NIR k-band continuum of NGC\,1052 obtained with NIFS. In white are shown optical continuum contours.}
	\label{fig2}
\end{figure}

In order to improve the quality of the data cube, we also applied the following treatment procedures to the data cube: correction of the differential atmospheric refraction, high spatial-frequency components removal with the Butterworth spatial filtering, ``instrumental fingerprint'' removal and Richardson-Lucy deconvolution. The final spatial resolution after the data treatment is 0\farcs79, estimated from a spatial profile obtained along the red wing of the broad H$\alpha$ emission, and the spectral coverage is 4317-6775\AA \citep{menezes+14,menezes+15}. The optical continuum of the treated datacube is shown on the left panel of Fig.~\ref{fig2}.

\subsection{Near-Infrared data}
The NIR data were obtained on October 2011 as part of GN-2010B-Q-25 Gemini North project using NIFS with ALTAIR adaptive optics system. Six 610\,s on target exposures were taken in the \textit{J} band and four 600 s exposures were taken in the \textit{K} band. The spatial resolution is 0\farcs1 and the spectral resolution is 6040 for the J band and 5290 for the K band.

\par

The data were reduced using the standard reduction scripts offered by the Gemini team, using tasks from the \textsc{Gemini/IRAF} package. The reduction included trimming of the images, flat fielding, sky subtraction, wavelength and s-distortion calibrations and telluric absorption removal. The flux calibration was performed by interpolating a blackbody function to the spectrum of the telluric standard star. The reduced datacubes were constructed with square spaxels of 0\farcs05$\times$0\farcs05. After the differential atmospheric refraction correction, the individual datacubes in each band were median combined to a single datacube. The combined datacubes were spatially re-sampled, in order to obtain spaxels of 0\farcs021. This spaxel size was chosen because it is a sub-multiple value of the original NIFS spaxel dimensions (0\farcs103$\times$0\farcs043). Such spatial re-sampling provides a better visualization of the contours of the structures. The Butterworth spatial filtering and the instrumental fingerprint removal were then applied. The final spectral coverage of the NIR data is 11472-13461\,\AA$ $ for the J band and 21060-24018\,\AA$ $ for the K band. 

\par 

\begin{figure}
	\includegraphics[width=\columnwidth]{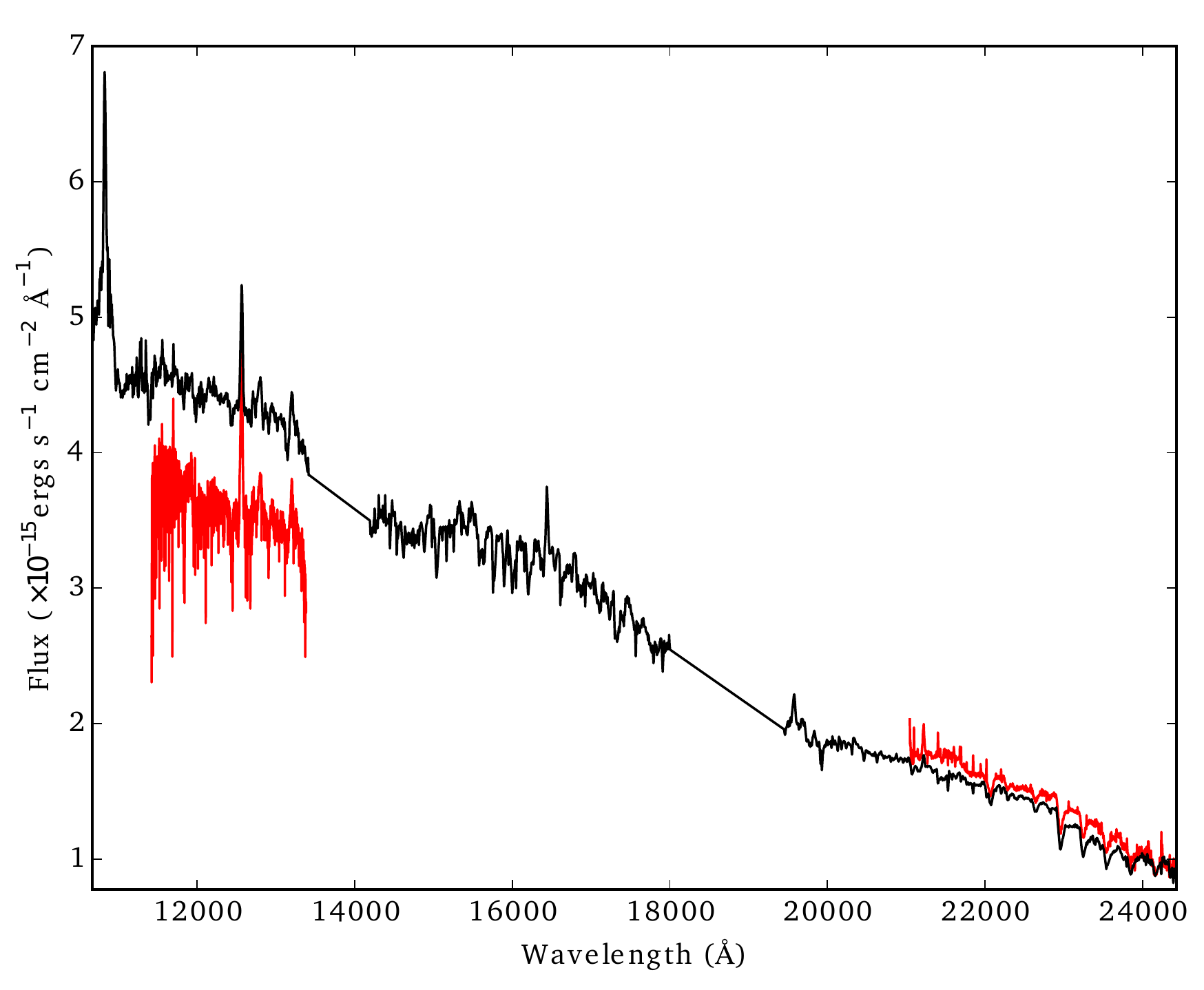}
	\caption{In black, GNIRS longslit spectra of NGC\,1052, as published by \citet{mason+15}. In red, we show the extracted NIFS spectra corresponding to the GNIRS aperture, both in J and K bands.}
	\label{scaling}
\end{figure}

In addition, the original NIFS field of view is 3\farcs0 x 3\farcs0 (320$\times$320\,pc$^2$), but due to low signal-to-noise ratio (S/N) on the borders of the FoV, we have shorten it to 2\farcs5$\times$2\farcs5. Since J and K bands were observed on different days, they may present some relative flux calibration problems. In order to minimize these effects, we have compared our data with a cross-dispersed spectrum of NGC\,1052 observed with Gemini near-infrared spectrometer (GNIRS) by \citet{mason+15}. This comparison was made using an extraction emulating the GNIRS aperture (slit oriented 90$^{\circ}$ east of north, with an aperture of 26$\times$155pc$^2$) in our datacubes. After that, we scaled the emulated NIFS long slit data with the GNIRS ones applying the obtained relation to all single spaxels of our data-cube. The GNIRS longslit spectra, together with the corresponding NIFS extracted spectra of the J and K bands are shown on Figure~\ref{scaling}. The average correlation of the J band was $\sim$80\%, whereas for the K band was $\sim$88\%. The NIR image of the k-band continuum is shown on right panel of Figure~\ref{fig2}.

\subsection{Panchromatic datacube}
\label{sec:PanCombine}

\begin{figure*}
  \includegraphics[width=\textwidth]{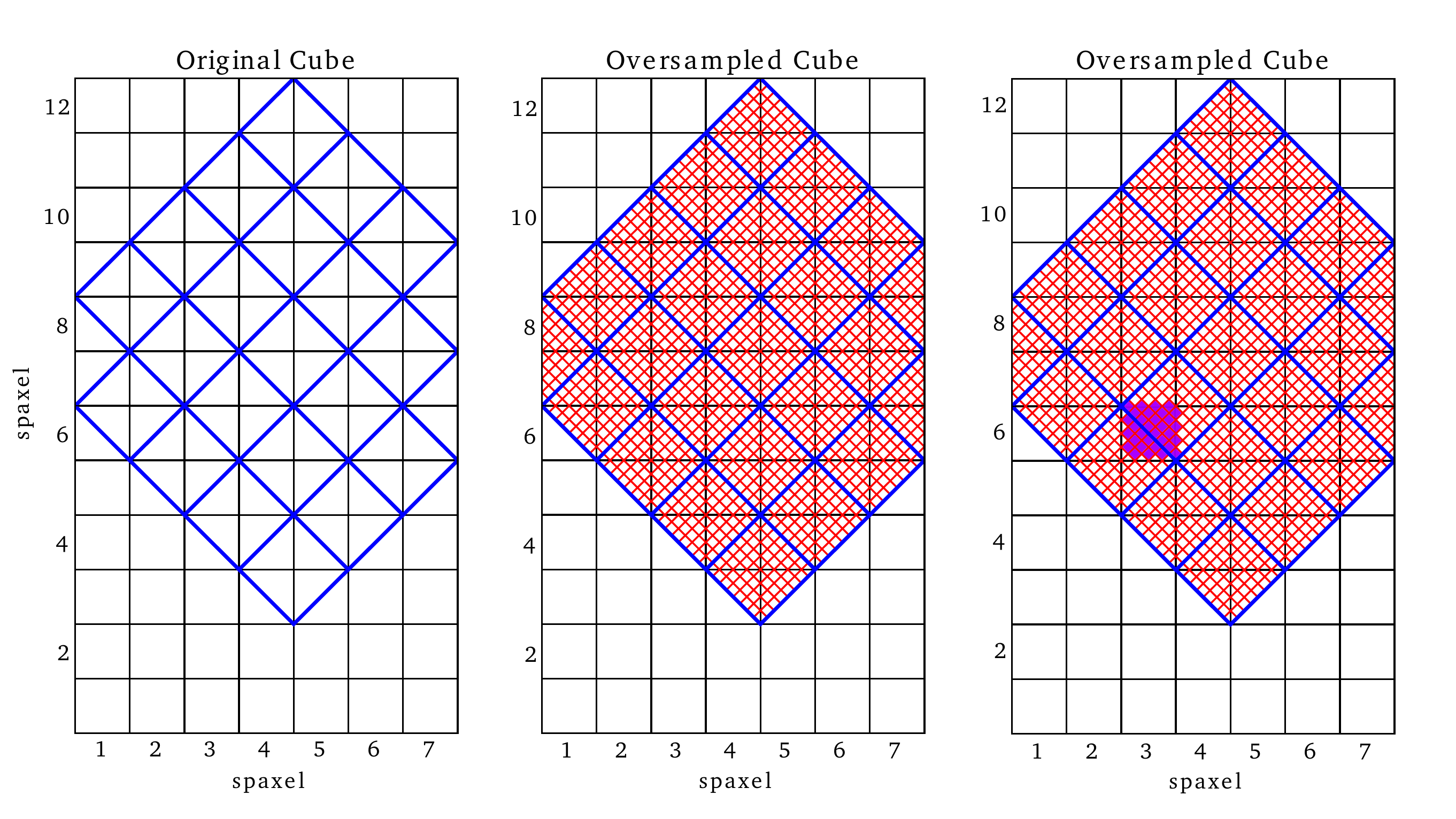}
  \caption{Combination of optical and NIR datacubes. In the left panel, optical data is represented by the black grid and NIR data is represented by the blue grid. In the middle panel, each NIR spaxel was sliced in 64 subspaxels (8$\times$8), represented by the red grid. In our data treatment, we sliced each spaxel in 900 subspaxels (30$\times$30) for better accuracy, rather than the 64 shown in the figure. In the right panel, all subspaxels marked in magenta were added to the optical (3,6) spaxel.}
  \label{fig:PanCombine}
\end{figure*}

\begin{figure*}
  \includegraphics[width=\textwidth]{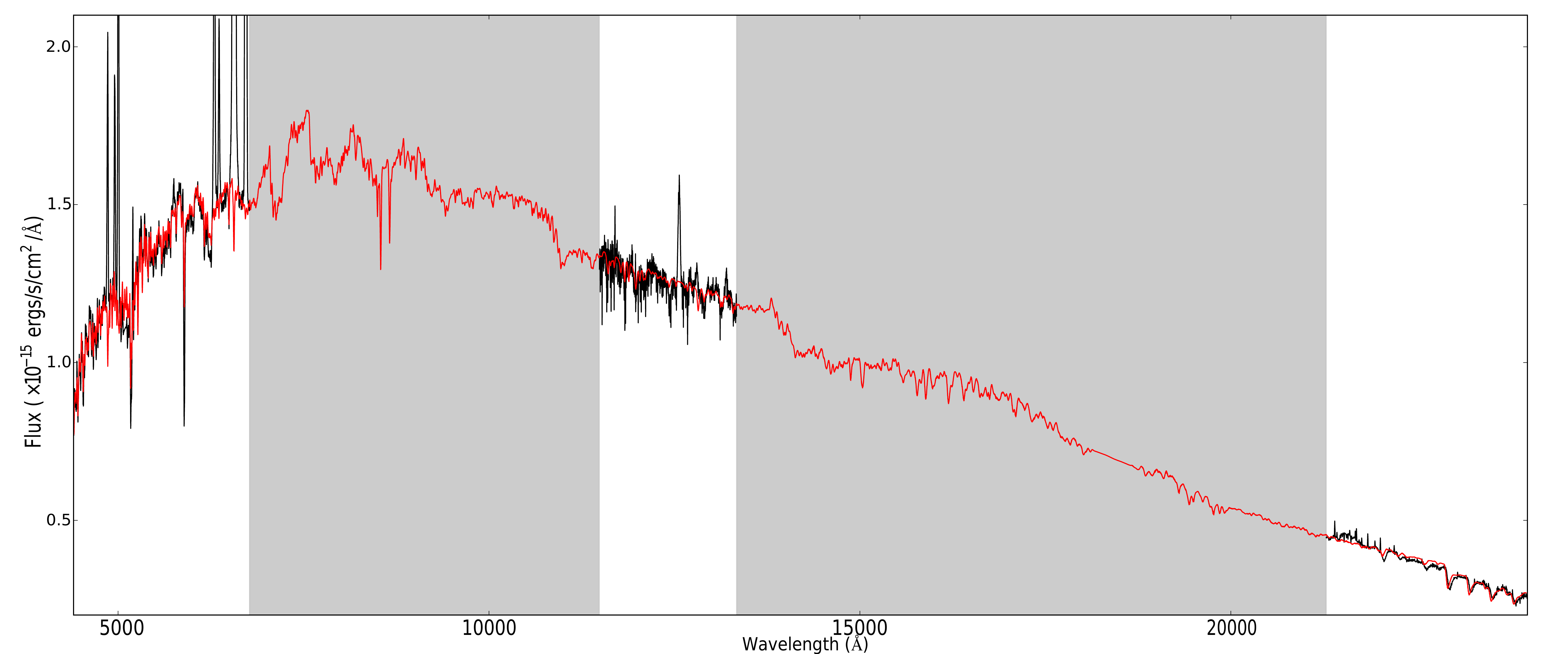}
  \caption{In black, we show the optical and NIR spectra of the combined datacube. Regions where we do not have data are grayed. In red, we present the corresponding spectrum modeled by {\sc starlight} with <t>$_L$=10.15 and <Z>$_L$=0.014.}
  \label{fig:PancFit}
\end{figure*}

For an improved characterization of the stellar population of the galaxy, we combined optical and NIR datacubes into a single panchromatic spatially resolved datacube. To do this, we followed the steps below:

\begin{enumerate}
  \item First, since the NIR datacube has a higher spatial resolution, we degraded it to match the optical data by convolving it with a gaussian of FWHM=0.7''. 
  \item Considering that optical and NIR data were observed on different nights and using different instruments, in order to correct the spectra for scale factors, we extracted an optical spectrum with the same spatial area of the \citet{mason+15} NIR longslit spectrum. The spectral separation between the end of the optical datacube and the beginning of the NIR longslit spectrum is small (<1700\AA), so that we searched for the scale factor that yielded the best agreement between the stellar population models and the combined optical+NIR spectrum. We did this by performing spectral synthesis on combined spectra with different scale factors. We found for this step an error $\leqslant$5\%. Lastly, we applied the scale factor to the NIR datacube.
    \item Assuming that the photometric center of NGC 1052 in both NIR and optical observations are located at the same position, we sliced each NIR datacube spaxel in 900 subspaxels (30x30) and added their fluxes to the optical spaxel that matched the central position of the NIR subspaxel. This is illustrated in Figure~\ref{fig:PanCombine}. The combined spectrum of the central spaxel and the spectrum modeled by {\sc starlight} are presented in Figure~\ref{fig:PancFit}.
\end{enumerate}

This approach of combining optical and NIR data might be useful to identify SSPs which are characterized by bluer colors, such as young populations, as well as stellar populations that are redder, such as intermediate-age ones. Also, since many important stellar features are located in the NIR \citep[like the 1.1$\mu m$ CN band, the 1.4$\mu m$ CN band and the ZrO features at 0.8-1.0$\mu m$, which are strong indicatives of TP-AGB stars and populations between 0.5 and 2\,Gyr][]{riffel+07,martins+13a,riffel+15,hennig+18}, adding this wavelength range to optical studies can further help to identify the stellar mixture of the galaxy.

\par

\section{Spectral Synthesis}
\label{sec:synth}

To study the stellar population of NGC 1052, we used the {\sc starlight} code \citep{CF04, CF05,fernandes+13}. Basically, the code fits the observed spectrum by a combination of SSPs in different proportions, considering reddening and line of sight velocity corrections. The final fit is carried out by {\sc starlight} searching for the $M_\lambda$ model that better describes the data. The model is given by

\begin{equation}
M_\lambda = M_{\lambda 0} [\sum_{j=1}^{N\star} x_j b_{j,\lambda} r_\lambda] \otimes G(v_\star, \sigma_\star),
\end{equation}

\noindent
where $M_{\lambda 0}$ is the flux at the normalization wavelength $\lambda 0$, $N_\star$ is the number of SSPs used to compose the model, $\vec x$ is the population vector so that $x_j$ indicates the contribution from the j-th SSP normalized at $\lambda 0$, $b_{j,\lambda}$ is the j-th model spectrum, $r_\lambda$ is the reddening term $r_\lambda=10^{-0.4(A_\lambda-A_{\lambda 0})}$, which is parameterized by the {\rm A$_V$}. The stellar velocity dispersions ($\sigma_\star$) and line-of-sight velocities ($v_\star$) are modeled by a Gaussian function $G(v_\star, \sigma_\star)$.a

\par

The best fit is carried out by minimizing the equation

\begin{equation}
\chi^2 = \sum_\lambda[(O_\lambda - M_\lambda)w_\lambda]^2,
\end{equation}

\noindent
where $O_\lambda$ is the observed spectrum. Emission lines and spurious features are masked out by setting $w_\lambda = 0$. Also, because of the lower S/N of the J band, we set $w_\lambda = 0.2$ to the whole wavelength range, both in the NIR and panchromatic synthesis. For the remaining spectral regions, we used $w_\lambda = 1.0$.

\par

To fit the spectra, we used the EPS models developed by \citet[][hereafter E-MILES]{e-miles}. These models are available with two possible evolutionary tracks and three possible initial mass functions (IMF). We chose the tracks of \citet{girardi+00} and the \citet{kroupa01} IMF, since in the NIR these models are based on the ones published by \citet{MIUSCATIR}, which we tested previously \citep{luisgdh+18} and found that they produce self-consistent results.
\par
These models were chosen because they cover the full spectral range of our data (3500 to 25000 \AA), offering a higher NIR spectral resolution when compared to other stellar population libraries \citep[e.g.][]{BC03, M05, C09}. This higher resolution is essential in order to produce better fits of the stellar absorptions.
\par
The only drawback of the E-MILES library is that it used the IRTF library, which does not include hot stars. As a consequence, in the NIR, only SSPs older than 500\,Myr are marked as safe by the authors. In the optical region, on the other hand, it includes SSPs as young as 30\,Myr. We were then able to test if there are no significant amounts of SSPs younger than 500\,Myr.
\par
Also, since NGC\,1052 is an elliptical galaxy, we used the reddening law from \citet{ccm} to reproduce dust extinction ({\rm A$_V$}), which is best suited for objects without active star formation.
\par
In order to remove the noise effects present in real data, a more consistent and robust way to present the stellar population is in the form of condensed population vectors \citep{CF04, CF05}. Following \citet{riffel+09}, and considering that the library of SSPs only includes models older than 1\,Gyr, we defined the light fraction population vectors as follows: xi(1\,Gyr $<$ t $\leqslant$ 2\,Gyr) and xo(t $>$ 2\,Gyr) to represent the intermediate-age and old stellar population vectors, respectively. 
\par
Also, to search for a possible emission from the AGN, we followed \citet{riffel+09} and added to the library of models a featureless continuum with $f_\nu \sim \nu^{-1.5}$. We also added to the NIR library Planck functions with temperatures between 700 and 1400\,K in order to reproduce the possible contribution from hot dust.
\par
Lastly, in order to better identify the stellar population mixture of the inner region of the galaxy, we followed \citet{CF05} and calculated the mean stellar age (\textit{<t>}) and mean metallicity (\textit{<Z>}). They are defined by the following equations:
\begin{equation}
<t>_L = \sum_{j=1}^{N\star}x_j log(t_j),
\end{equation}
\begin{equation}
<Z>_L = \sum_{j=1}^{N\star}x_j Z_j,
\end{equation}

where $t_j$ and $Z_j$ are the age and metallicity of the j-th SSP. The $x_j$ percentage contribution can be weighted by light (L) and mass (M) fractions. We normalized optical and panchromatic data at 4500\AA, and NIR data at 21910\AA. We also used these values to calculate the luminosity contribution. These values were chosen because they contain relatively few stellar absorptions and good S/N.

\section{Results}
\label{sec:results}

\subsection{Stellar Synthesis}
\label{sec:StellarSynthesis}

\begin{figure*}
  \includegraphics[width=\textwidth]{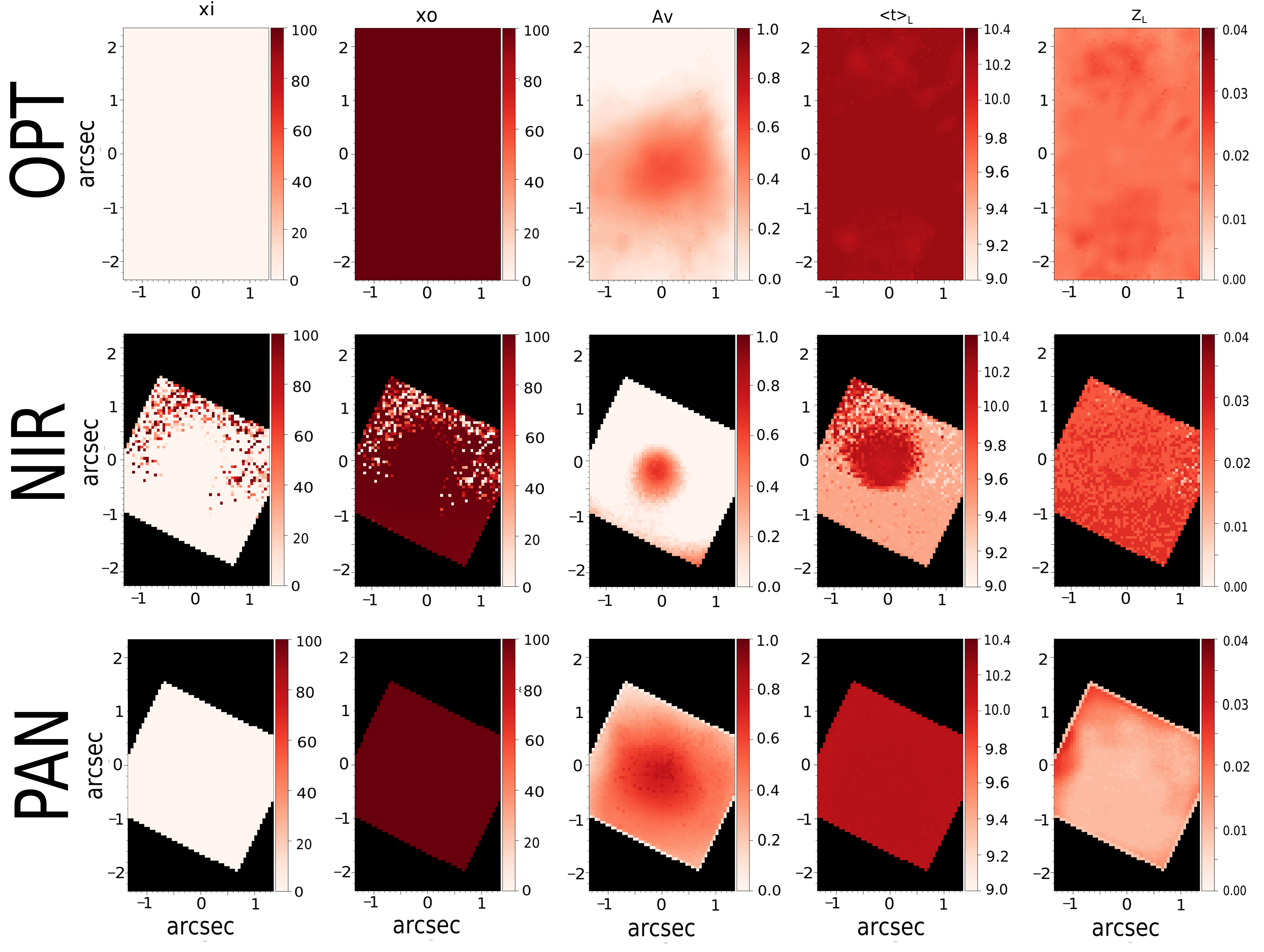}
  \caption{Synthesis results for NGC\,1052. From left to right, we present the spatial distribution of intermediate-age and old stellar populations, the {\rm A$_V$}, $<t>_L$ and lastly the $Z_L$. The top five panels show the optical results, the middle panels present the NIR results and the bottom panels show the results obtained with the combination of optical and NIR data.}
  \label{fig:StPop}
\end{figure*}

By fitting the optical datacube with E-MILES library, we found only old stellar content in the entire datacube. Looking at the <t>$_L$, it is possible to see that the cube is dominated by a stellar population of $\sim$12\,Gyr, with some locations displaying a slightly younger (t$\sim$10\,Gyr) stellar population. Also, no contribution from a featureless continuum was found. The {\rm A$_V$} throughout the cube was very low ($\sim$0.2\,mag), with the nucleus peaking at {\rm A$_V$}=0.5\,mag. We found metallicity values close to solar \citep[Z$_\odot$=0.019,][]{girardi+00} in the entire cube. We show in the top panels of Figure~\ref{fig:StPop} the spatial distribution of intermediate-age and old stellar populations, as well as the {\rm A$_V$}, $<t>_L$ and $Z_L$ for the optical results.

When fitting the NIR datacube, we found the same results in the nucleus of the galaxy, with a dominance of very old (t$>$10\,Gyr) stellar populations, a higher dust reddening ($\sim$0.8\,mag) and solar metallicity. Because of the better spatial resolution of NIR data, the reddening peak is much more concentrated and intense compared to the optical one. Also, no contribution from Planck functions nor from a featureless continuum were found.
\par
However, according to the NIR data, the circumnuclear population is dominated by a $\sim$2.5\,Gyr population, which contributes to $\gtrsim$80\,\% of the light. Although this population is classified as old (t>2\,Gyr) after binning the results, there is a clear circumnuclear drop in the <t>$_L$ values. The reddening in this region is much lower than the nucleus ($<$0.2\,mag), and there is no significant difference in the metallicity if compared to the nucleus. We present these results in the middle panels of Figure~\ref{fig:StPop}.
\par
By performing the synthesis using the panchromatic datacube, we found again a dominance of old stellar populations. No difference in the stellar population can be seen in the <t>$_L$ panel, with the whole panchromatic region displaying a $\sim$10\,Gyr stellar population. Also, we found a reddening contribution compatible with both optical and NIR results, as well as a metallicity slightly lower (Z$\sim$0.010) than solar throughout the datacube. The only exceptions were the borders of the cube, which are affected by a low S/N. As in optical and NIR results, we did not detect contributions from either Planck functions or a featureless continuum. The results for the combined datacube are presented in the bottom panels of Figure~\ref{fig:StPop}.
\par
Since the panchromatic synthesis favored optical results, we performed the synthesis again, increasing the NIR weight. By setting $w_{NIR}=2$, the results from {\sc starlight} are the same, with a dominance of old stellar populations throughout the cube. When setting $w_{NIR}=4$, the results found lie in between optical and NIR ones, with a dominance of old stellar populations in the datacube but with <t>$_L\sim$8\,Gyr at the borders of the FoV. However, for all fits with $w_{NIR}>2$, the fits are not able to match the optical spectral region.

\subsection{Absorption Band Measurements}
\label{sec:AbsBands}

In order to avoid contamination due to stellar kinematics when measuring the equivalent widths (EW) of the absorption bands, we first corrected the spectra for Doppler shift, using the line-of-sight velocities derived in Section~\ref{sec:StellarKinematics}. We then applied a python version of the pacce code \citep{pacce} for each spaxel and measured the EWs based on the line limits and continuum bandpasses of \citet{worthey+94} for the optical region and of Riffel et al. (in prep.) for the NIR, which are presented in Table~\ref{tab:EW}. 

\begin{figure*}
  \includegraphics[width=\textwidth]{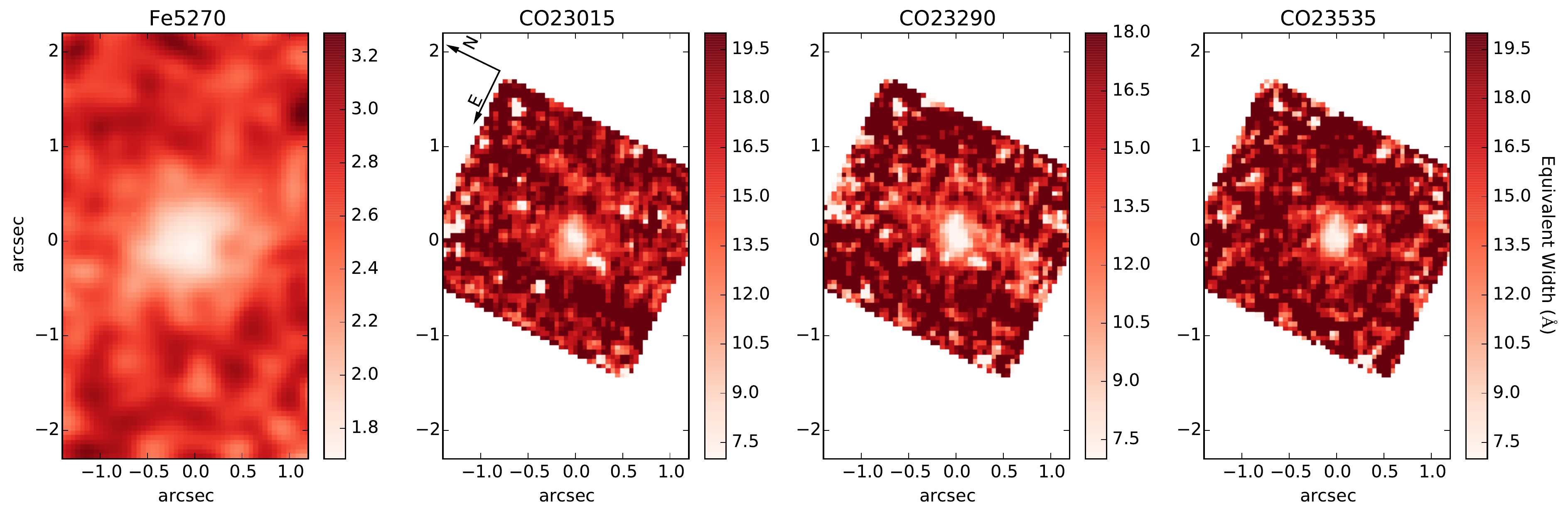}
  \caption{Equivalent width spatial distribution of the optical Fe$\lambda$5265\AA and the NIR CO bands.}
  \label{fig:EWmaps}
\end{figure*}

From the optical bands, only Fe$\lambda$5270\AA$ $ was not contaminated by emission from the gas and with enough S/N. The NaD$\lambda$5895\AA$ $ absorption band, although being the strongest in the optical region, was contaminated by the HeI$\lambda$5875\AA$ $ emission line, as well as by the interstellar sodium. Also, the Mg$_2\lambda$5175\AA$ $ band was contaminated by the [NII]$\lambda$5198\AA$ $ emission line. In the NIR, on the other hand, the three CO bands near 23000\AA$ $ were isolated and with a good S/N ratio ($\sim$100 for the nucleus). The EW maps for the optical FeI$\lambda$5270\AA$ $ and the three CO bands in the NIR are shown in Figure~\ref{fig:EWmaps}. 

\begin{table*}
  \centering
  \setlength{\tabcolsep}{2pt}
  \caption{Line limits and continuum bandpasses. \label{tab:EW}}
  \begin{tabular}{cccc}
  \hline
  \noalign{\smallskip}
  Centre &     Main        & line limits   & continuum bandpass \\
   (\AA) &   Absorber    & (\AA)         &  (\AA) \\
  \hline
  \noalign{\smallskip}            
  5265.65  &    Fe                     &   5245.650-5285.650 &    5233.150-5248.150, 5285.650-5318.150   \\
  23015.0 &    CO                     & 22870.000-23160.000 &   22700.000-22790.000, 23655.000-23680.000\\
  23290.0 &    CO                     & 23160.000-23420.000 &   22700.000-22790.000, 23655.000-23680.000\\
  23535.0 &    CO                     & 23420.000-23650.000 &   22700.000-22790.000, 23655.000-23680.000\\
  \hline
  \end{tabular}
  \begin{list}{Table Notes:}
  \item The optical indices are based on \citet{worthey+94} and the NIR indexes are based on Riffel et al. (in prep).
  \end{list}
\end{table*}

The Fe5270 map shows a constant 2.76$\pm$0.17$ $\AA$ $ EW throughout the cube, with lower ($\sim$1.8\,\AA) values closer to the nucleus with a 0\farcs6 FWHM. The three CO absorptions located in the end of the K band present the exact same spatial distribution, with a constant ($\sim$18\,\AA) EW throughout the cube and a lower ($\sim$10\,\AA) EW in the center. Because of the higher spatial resolution, the EW drop as seen in the NIR is much more concentrated, with a FWHM of 0.26. These results suggest that the component that causes this drop is emitted very close to the nucleus of the galaxy.

\subsection{Stellar Kinematics}
\label{sec:StellarKinematics}

To derive the stellar kinematics, we used the penalized pixel-fitting (pPXF) code, described in \citet{ppxf}. Like {\sc starlight}, this code also searches for the combination of spectra from a user-provided library to fit the model that better reproduces the observed spectrum. We chose to perform the fitting of the stellar kinematics with pPXF because it allows the user to remove continuum information with a high-order polynomial, so that more weight is given to the stellar features.
\par
In the optical region, we used the \citet{e-miles} stellar library, since it has an adequate spectral resolution and coverage. For this wavelength range, we fitted the whole cube, masking only the emission lines. For the NIR stellar templates, on the other hand, we used the \citet{winge+09} library of late spectral type stellar templates, which was observed with GNIRS, and provides a better spectral resolution (R>5000) and S/N, being limited to the end of the K-band (21500<$\lambda$<24200\,\AA). Since NIR data has a smaller (S/N), especially the K-band, we limited the fitting region to 22830-24025\,\AA, where the CO bands are present.
\par
\begin{figure*}
  \includegraphics[width=\textwidth]{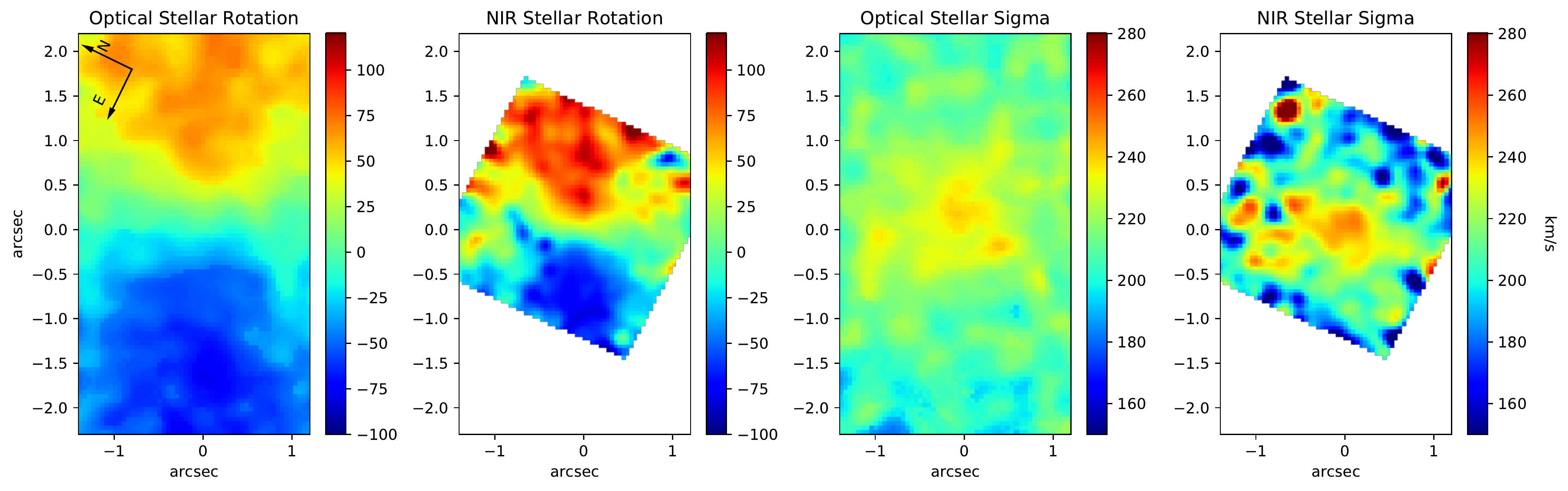}
  \caption{Optical and NIR stellar kinematics derived by pPXF}
  \label{fig:StellarKinematics}
\end{figure*}

The pPXF code returns as output values for the radial velocity ($v_c$) and stellar velocity dispersion ($\sigma$) for each spatial position. The results are presented in Figure~\ref{fig:StellarKinematics}.

In order to search for deviations from circular motions, we followed \citet{bertola+91} and assumed that the stars are on circular orbits in a plane with a rotation curve given by

\begin{equation} 
v_c = \frac{Ar}{(r^2 + c_0^2)^{p/2}},
\end{equation} 
\noindent
where $A$, $c_0$ and $p$ are parameters to be found and $r$ is the radius. Then, the observed radial velocity at a position ($R$,$\Psi$) on the plane of the sky is given by:

\scriptsize
\begin{equation}
v(R, \Psi) = v_{sys} + \frac{AR cos(\Psi-\Psi_0)sin\theta cos^p \theta}{\{R^2[sin^2(\Psi-\Psi_0)+cos^2\Theta cos^2(\Psi-\Psi_0)]+c_0^2 cos^2\Theta  \}^{p/2}},
\end{equation}
\normalsize
\noindent
where $\Theta$ is the disk inclination ($\Theta$=0 being is a face-on disk), $\Psi_0$ is the position angle of the line of nodes and $v_{sys}$ is systematic velocity of the galaxy. We then developed a script that automatically searches for the center, inclination and velocity amplitude by performing a Levenberg-Marquardt $\chi^2$ minimization.
\par
\begin{figure*}
  \includegraphics[width=\textwidth]{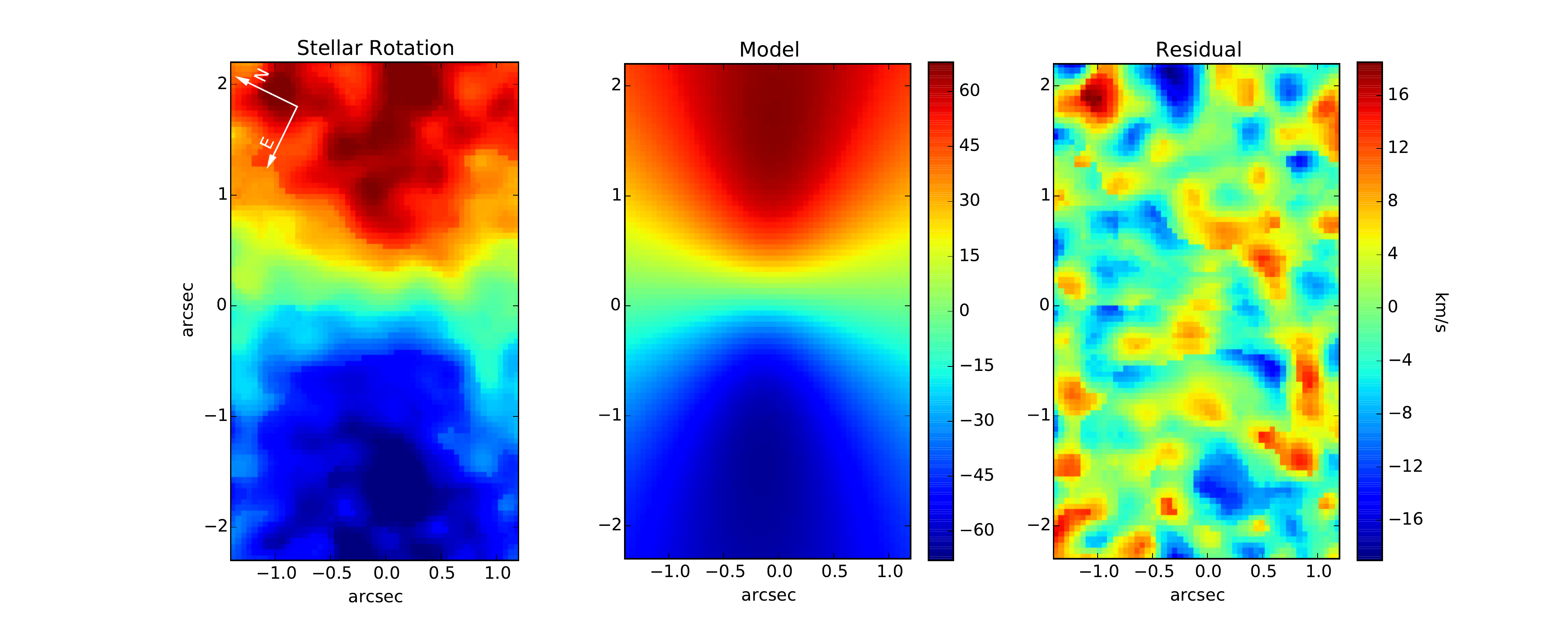}
  \caption{Stellar rotation, model and residual for the optical data.}
  \label{fig:StellarKinematics2}
\end{figure*}

Because of the lower S/N of the NIR data, we fitted only the rotation obtained with  with optical data. In Figure~\ref{fig:StellarKinematics2}, we present the stellar velocity field, the best fit model, and the residual map, obtained by the subtraction of the model from the observed velocities. The residual is, in all locations, below $|16|$ km$\cdot$s$^{-1}$.
\par
These results show that the stellar motions in the central $\sim$500\,pc of NGC\,1052 are dominated by two components:  random orbits in the bulge with $\sigma\sim$240\,km\,s$^{-1}$ and circular orbits in the galaxy plane with speeds up to 100\,km\,s$^{-1}$.

\section{Discussion}
\label{sec:discussion}

\subsection{Divergence in optical and NIR results}
\label{divergence}

The optical results for the stellar population synthesis, revealing that this galaxy is dominated by old stellar content in the nuclear region, are in agreement with past results, which showed that NGC\,1052 is dominated by stellar populations older than 2\,Gyr \citep{raimann+01,milone+07,pierce+05}. Even though one previous paper about this galaxy found spots suggesting younger globular clusters \citep[t$\sim$7\,Myr][]{ontiveros+11}, these regions are located outside the FoV of both the optical and NIR datacubes.
\par
On the other hand, NIR results revealed a circumnuclear ring of $\sim$2.5\,Gyr (contributing up to 100\% of the light in some regions). This stellar population can be real, in which case NIR data was able to access stellar populations which were not found through optical data. By using NIR data, similar circumnuclear rings of intermediate-age populations were already reported before \citep{riffelRA+10,riffel+11,storchi-bergmann+12,diniz+17}, suggesting that this population can be real. However, the lack of any contribution from stellar populations younger than 10\,Gyr in the optical spectra suggests a different explanation, in which these populations are being mistakenly identified by {\sc starlight} as a consequence of other factors, discussed below.

\subsection{Panchromatic synthesis}
\label{panchromatic_discussion}

The results obtained by combining optical and NIR data favor optical results, with a dominance of old ($\sim$10\,Gyr) stellar populations in the entire FoV. Also, NIR data is well reproduced by the panchromatic fit, suggesting that the result is degenerated. Lastly, our NIR results are in contrast with literature results, which show that elliptical galaxies are dominated by old stellar populations \citep[e.g.][]{PadillaStrauss08,zhu+10}.
\par
Knowing that the intermediate-age circumnuclear ring found in the NIR completely vanished after combining optical and NIR data (which added information allowing the identification of a broader range of stellar populations), our panchromatic results suggest that other factors might me playing an important role in the determination of the stellar population of the NIR data, besides the stellar properties of the galaxy itself. First, we do not have H-band data, which contains more stellar absorptions when compared to J and K-bands. Second, the low signal-to-noise ratio of the J precludes the presence of enough constraints to {\sc starlight}, besides the shape of the continuum. As reported in the literature, the fit of the absorption bands is crucial to the correct determination of the stellar population of the galaxy using NIR data \citep{baldwin+18,luisgdh+18}. These results combined suggest that the circumnuclear ring found in the NIR might be biased towards younger ages.
\par
Since optical results detected a dominance of old stellar populations, and considering the elliptical nature of NGC\,1052, we did not expect to detect a different stellar population by adding NIR data. However, for galaxies dominated by intermediate-age stellar populations \citep[which display unique NIR features,][]{M05, riffel+07, riffel+09}, this technique of adding NIR data to optical studies might be useful in order to unambiguously determine their stellar populations. This technique might also be useful to study the contents of more luminous AGNs, such as Seyfert galaxies, due to the usual  presence of a strong featureless continuum, emitted by the AGN, which peaks in the optical or even bluer wavelengths. The NIR also allows the constraint of the contribution of the AGN hot nuclear torus.

\subsection{Equivalent Width Drop}
\label{EWdrop}

For the optical Fe5270 index, we compared the EW values with the ones available online for the E-MILES library. For SSPs younger than 1\,Gyr, their derived values for this absorption are much smaller than ours, with values close to zero for the younger models and subsolar metallicity. Our off-center measurements have an average value of 2.76$\pm$0.17$ $\AA, and are compatible with a wide range of values. For solar metallicity, the compatible ages range from 2.5 to 6.5\,Gyr, whereas for subsolar metallicity the range goes from 1.0 to 4.0\,Gyr and for supersolar metallicities range from 6.5\,Gyr up to the older models with 14.0\,Gyr.

\par
In order to compare NIR absorptions, due to the lack of measurements in the literature, we employed our technique to the E-MILES models and measured their equivalent widths. For the safe models available (t>0.5\,Gyr), the EW values range from 14.0 to 18.5\,\AA$ $ for the first CO absorption, 14.4 to 18.8\,\AA$ $ for the second, and 15.4 to 21.1\,\AA$ $ for the third. These measurements are compatible with all our extranuclear measurements for these bands within the standard error of the data (average EW of 17.5$\pm$1.2, 17.8$\pm$1.0, 20.4$\pm$1.1 for the three CO bands respectively). However, no valuable information can be extracted from these absorptions concerning the stellar population of the galaxy, since nearly all SSPs from 0.5\,Gyr to 14\,Gyr match our data properly.
\par
The spatial distribution of the stellar absorptions shows a nuclear EW drop in the nuclear region, both in the optical and in the NIR, with the same spatial profile of the PSF of the data. These results suggest that the region in which the EW drop is produced is not spatially resolved, being confined to the region dominated by the LLAGN.
\par
Since the spectral synthesis did not detect any contribution from a featureless continuum or a nuclear variation in the stellar populations, these results indicate that a stellar population synthesis is unable to detect all components that contribute to the emission of this galaxy. It is worth noting that, even the optical synthesis, which was performed with E-MILES library and included both young SSPs and a featureless continuum, did not detect such components.
\par
There are two possible explanations for these EW drops. First, this can be caused by a difference in the stellar population in the central region, such as a young stellar population in which only the hydrogen absorptions are present, and thus could dilute the EWs of the other absorption lines. A second explanation is a featureless continuum caused by the AGN, which can dilute the stellar population features.
\par
Since this galaxy is elliptical, suggesting that high contributions from young populations are not likely to occur, and since it hosts a LLAGN \citep{wrobel84,FeyCharlot97,kadler+04,barth+99,sugai+05}, this nuclear EW drop points toward a featureless continuum associated with the LLAGN.
\par
A similar result was already reported by \citet{burtscher+15}, which analyzed 51 local AGNs, with NGC\,1052 as part of the sample, and found the same EW nuclear drop. They also associated this drop with the contribution from a featureless continuum from the AGN hosted by this galaxy.

\subsection{Stellar Kinematics}
\label{Kinematics_discussion}

The kinematic results, showing that this galaxy displays a circular rotation and a broadened nuclear dispersion, are in agreement with the ones derived by \citet{dopita+15} and \citet{riffelRA+17}. The stellar kinematics of this galaxy is characterized by a rotation with a major axis oriented along the East-West direction (slightly inclined toward Southeast-Northwest) with blueshifted velocities in the eastern region and redshifted velocities in the western region. Also, the stellar sigma peaks at $\sim$240\,km$\cdot$s$^{-1}$ in the nucleus, with lower ($\sim$200\,km$\cdot$s$^{-1}$) values closer to the borders. In the NIR, even with a lower S/N, it is possible to see that the rotation pattern agrees with that seen in the optical. The dispersion derived from the NIR data, on the other hand, is $\sim$20km$\cdot$s$^{-1}$ higher if compared to optical values, which is probably a feature caused by the lower S/N of the NIR data.
\par
The stellar sigma peak is cospatial with the EW drop in the absorption lines, suggesting that these two phenomena might be related. The most probable explanation is that both of them are related to the nuclear region, with the sigma peak being also caused by the sphere of influence of the SMBH.

\section{Conclusions}
\label{sec:conclusions}

In this work, we studied the spatial distribution of the stellar spectral properties of the inner 320$\times$535\,pc$^2$ of NGC\,1052, both in the optical and in the NIR. Our results can be summarized as follows:
\par
The stellar population in the optical is dominated by old (t$>$10\,Gyr) components, with low ($\sim$0.5) nuclear dust extinction and solar metallicity throughout the galaxy. When performing the synthesis in the NIR, on the other hand, a circumnuclear ring of stellar populations with ages $\sim$2.5\,Gyr was found, also with low ($\sim$0.8) nuclear dust extinction and solar metallicity throughout the galaxy.
\par
When performing the synthesis in the combined optical and NIR datacube, we found results compatible with the ones found in the optical, with a dominance of old (t$>$10\,Gyr) stellar populations with low ($\sim$0.5) nuclear dust extinction and solar metallicity in the entire FoV. 
\par
The absorption bands, both in the optical and in the NIR, display a drop in the equivalent widths in the nucleus. This drop was not detected in the stellar population synthesis, indicating that a stellar population synthesis is unable to identify all components that contribute to the emission of this galaxy. We find that the presence of a featureless continuum emitted by the LLAGN is the most plausible scenario to explain these drops.
\par
Lastly, the stellar kinematics are dominated by two components. Random motions, as shown by the high ($\sim$240\,km$\cdot$s$^{-1}$) velocity dispersion, dominate the nucleus, with the remaining regions dominated by stars rotating in a plane around the center with speeds up to 100\,km$\cdot$s$^{-1}$.

\section*{Acknowledgements}

We thank the anonymous referee for reading the paper carefully and providing thoughtful comments that helped improving the quality of the paper. LGDH thanks CNPq. RR thanks CNPq and FAPERGS for financial support. NZD thanks to CNPq for partial funding. RAR thanks support from CNPq and FAPERGS. Based on observations obtained at the Gemini Observatory and processed using the Gemini IRAF package, which is operated by the Association of Universities for Research in Astronomy, Inc., under a cooperative agreement with the NSF on behalf of the Gemini partnership: the National Science Foundation (United States), the National Research Council (Canada), CONICYT (Chile), Ministerio de Ciencia, Tecnolog\'ia e Innovaci\'on Productiva (Argentina), and Minist\'erio da Ci\^encia, Tecnologia e Inova\c{c}\~ao (Brazil). This study was financed in part by the Coordena\c{c}\~ao de Aperfei\c{c}oamento de Pessoal de N\'ivel Superior - Brasil (CAPES) - Finance Code 001.


\bibliographystyle{mnras}
\bibliography{luisgdh} 



\appendix


\bsp	
\label{lastpage}
\end{document}